\documentclass[twocolumn]{aastex62}
\usepackage{natbib}
\usepackage{amsmath}
\usepackage{afterpage}

\begin{document}

\title{Longitudinally Modulated Dynamo Action in Simulated M-Dwarf Stars}
\author{C.P. Bice}
\affiliation{JILA and Department of Astrophysical and Planetary Sciences, University of Colorado Boulder}
\author{J. Toomre}
\affiliation{JILA and Department of Astrophysical and Planetary Sciences, University of Colorado Boulder}

\correspondingauthor{Bice, Connor P.}
\email{connor.bice@colorado.edu}

\begin{abstract}
M-dwarf stars are well known for the intense magnetic activity that many of them exhibit. In cool stars with near-surface convection zones, this magnetic activity is thought to be driven largely by the interplay of convection and the large scale differential rotation and circulations it establishes. The highly nonlinear nature of these flows yields a fascinatingly sensitive and diverse parameter space, with a wide range of possible dynamics. We report here on a set of three global MHD simulations of rapidly rotating M2 (0.4 $M_\odot$) stars. Each of these three models established \it nests of vigorous convection \rm that were highly modulated in longitude at low latitudes. Slight differences in their magnetic parameters led each model to disparate dynamo states, but the effect of the convective nest was a unifying feature. In each case, the action of longitudinally modulated convection led to localized (and in one case, global) reversals of the toroidal magnetic field, as well as the formation of an active longitude, with enhanced poloidal field amplitudes and flux emergence. 
\end{abstract}


\section{Introduction}
Despite their small sizes, cool temperatures, and dim luminosities, M-dwarf stars are well known for the vigorous magnetism many of them display. Unlike more massive stars, of which less than 20$\%$ demonstrate chromospheric markers for magnetic activity, nearly all fully convective (FC) late M-dwarfs appear to be highly magnetically active (e.g. \citealt{West08}; 2015). This transition in activity occurs sharply over a range of stellar masses centered on 0.35 M$_\odot$, below which main sequence stars are FC, and has come to be called the tachocline divide. As the convective stability and abundant shear of a tachocline is often held to be a crucial ingredient in the dynamos of Sun-like stars , its absence in FC stars is speculated to necessitate fundamentally different dynamo action, which may explain the difference of activity level. However, recent measurements of activity on slowly rotating FC stars \citep{Wright18} have found that there is no significant difference in their rotation-activity relations compared to more massive stars, suggesting that the underlying dynamo processes may be shared.

On chromospherically active stars, flares are exceedingly common events \citep{Kowalski09}. Some M-dwarfs appear to have surfaces nearly carpeted by magnetic fields at strengths similar to sunspots ($\sim 10^3$ G), and give off flares that may be a thousand times more energetic than those of the Sun (e.g. \citealt{Kowalski10}; \citealt{Silverberg16}; \citealt{Davenport16}). The magnetic fields responsible for this activity at the surface of a star must inevitably be tied to that star's internal dynamo. As our ability to make direct measurements of the properties of flows and magnetic fields beneath a star's photosphere is limited, the study of dynamo action in stars is largely the domain of theory and computation.

\subsection{Convective Dynamo Simulations}
Solar convection and dynamo theory has made substantial advances through 3-D global simulations carried out in spherical geometry (e.g. \citealt{brun17}; \citealt{charbonneau20}). Early work showed that turbulent convection influenced by rotation can build strong magnetic fields within the solar convection zone (CZ) itself (e.g. \citealt{brun04}), while yielding a differential rotation profile in latitude of fast equator and slower poles in reasonable accord with helioseismic findings. The discovery of the solar tachocline led to suggestions that an \it interface dynamo \rm exploiting that shear may be crucial to the observed solar magnetic cycling with the emergence of strong fields as sunspots. More recent dynamo models with the inclusion of a tachocline of shear at the base of the modeled CZ (e.g. \citealt{browning06}; \citealt{ghizaru10}; \citealt{passos14}; \citealt{augustson15}; \citealt{strugarek18}; \citealt{matilsky19}) revealed that the building of stronger mean magnetic fields is so favored, and that the cycling periods were typically longer.  A separate realization was that dynamo action within the CZ itself, with increasing levels of turbulence and rotational constraint, could build strong wreaths of toroidal magnetism that appeared as coherent structures, and that these could be capable of periodic cycles (e.g. \citealt{Glatzmaier85}; \citealt{brown10}; \citealt{brown11}; \citealt{matilsky20}).

Relatively few such global convective dynamo studies have been conducted in the domain of M-dwarf stars. Early work by \citet{browning08} considering lower mass FC M-dwarfs found that the deep CZ could support very strong non-axisymmetric fields, which strongly quenched the star's differential rotation. Later, more turbulent simulations of FC M-dwarf stars led to a number of interesting results. \citet{yadav15a}a found strong, axisymmetric fields which were statistically steady in time and recovered many of the observed characteristics of M-dwarf surface fields. A somewhat slower rotating model \citep{yadav15b}b revealed that flux concentration by merging downflow lanes could lead to the formation of large, persistent high-latitude starspots in these stars. A still slower rotating model \citep{yadav16} built large-scale, axisymmetric, cycling magnetic fields of somewhat lower amplitude which did not eliminate the star's differential rotation, reminiscent of the distributed $\alpha\Omega$ type dynamos prevalent in solar-like contexts. In \citet{bice20} (hereafter BT20), we presented an exploration of the influence exerted by a tachocline in more massive, shell-convecting M-dwarf stars as a contributing factor to the break in observed magnetic activity across the tachocline divide. Our models produced a wide variety of field configurations, nearly all of which led to quenching of the differential rotation to a significant degree. We found that including a tachocline in models of early M-dwarf stars led to their surface fields being more favorable for rapid stellar spin-down, which may contribute to the formation of the tachocline divide.

Although convective dynamo simulations represent the best tool we have for studying the generation of magnetic fields within stars, they have struggled to robustly capture the formation and rise of magnetic flux tubes, which is thought to be the mechanism underlying the creation of sunspots and starspots. Nevertheless, as access to computational resources continues to expand, bringing with it ever more turbulent parameter spaces, we are beginning to see flux emergence as a general feature of these models (e.g. \citealt{nelson11}; \citealt{nelson13}; \citealt{nelson14}; \citealt{fan14}). The flows and fields of simulated stellar CZs which did not themselves generate buoyant flux ropes have been used as backgrounds against which to study how thin flux tubes \it might \rm have risen. \citet{weber16} found that in FC M-dwarfs, inserted flux tubes preferred to rise parallel to the axis of rotation, strongly favoring mid to high latitudes as emergence locations. The surface fields of simulated M-dwarfs are beginning to inform studies of activity in their atmospheres (e.g. \citealt{alvarado19}).

The work presented here largely concerns the interactions between a \it nest of longitudinally modulated convection \rm and the magnetic fields induced both by that convection and within the stellar tachocline. Convective nests have also been observed and studied in fast-rotating solar-like contexts far from convective onset with $\mathrm{P_r}=1/4$ \citep{brown08}. There, the degree of longitudinal modulation appeared to be tied to the rotation rate, with faster rotation leading to convection that was more confined to the nest. The nests propagated prograde along the equator, and spanned the full height of the CZ, remaining coherent over hundreds of rotation periods. In the most strongly modulated models, low-latitude convection outside of the nests was almost entirely suppressed, with gently streaming zonal flows connecting the nest's trailing edge to its front. The authors proposed that these nests may be contributing to the magnetic active longitudes that have been observed on the Sun and other stars (e.g. \citealt{bogart82}; \citealt{bumba90}; \citealt{berdyugina02}). Longitudinally modulated convection has also been noted as traveling waves in laboratory experiments and experiments studying binary-fluid convection (\citealt{walden85}; \citealt{moses86}; \citealt{heinrichs87}) and thermosolutal convection (\citealt{deane88}; \citealt{spina98}).

\section{Formulating the Problem}

For the simulations of M2 (0.4 M$_\odot$) stars presented here, we employ the open-source 3D MHD code Rayleigh \citep{rayleigh} to evolve the anelastic compressible equations in rotating spherical shells. Rayleigh is a pseudospectral code, employing both a physical grid and a basis of spherical harmonics and Chebyshev polynomials. The anelastic equations are a fully nonlinear form of the fluid equations from which sound waves have been filtered out. This provides an appropriate framework for exploring subsonic convection within stellar interiors, where fast-moving p-modes would otherwise throttle the maximum timestep. The thermodynamic variables are linearized against a one-dimensional, time independent background state involving density, pressure, temperature, and entropy ($\bar{\rho},\,\bar{P},\,\bar{T},$ and $\bar{S}$, respectively), with deviations from the background written without overbars. As with all simulations of this type, the viscosity $\nu$, conductivity $\kappa$, and resistivity $\eta$ are inflated by many orders of magnitude as a parameterization of the turbulent mixing occurring at sub-grid scales. The detailed forms of the anelastic equations involving the velocity vector $\mathbf{v}$ and the magnetic field vector $\mathbf{B}$ solved in Rayleigh are as follows.

\begin{equation}
\begin{split}
\mathrm{Mome}&\mathrm{ntum:}\;\;\bar{\rho}(\frac{D\mathbf{v}}{Dt}+2\Omega_0\hat{z}\times\mathbf{v})=\\&-\bar{\rho}\nabla\frac{P}{\bar{\rho}}+ \frac{\bar{\rho}g}{c_p}S+\nabla\cdot\mathcal{D}+\frac{1}{4\pi}(\nabla\times\mathbf{B})\times\mathbf{B}\;,
\end{split}
\label{eqn:momentum}
\end{equation}
\begin{equation}
\begin{split}
\mathrm{Energy:}\;\;\bar{\rho}\bar{T}\frac{DS}{Dt}&=\nabla\cdot[\kappa\bar{\rho}\bar{T}\nabla S]+\\&Q+\Phi+\frac{\eta}{4\pi}[\nabla\times\mathbf{B}]^2,
\end{split}
\label{eqn:thermal}
\end{equation}
\begin{equation}
\mathrm{Induction:}\;\;\frac{\partial\mathbf{B}}{\partial t}=\nabla\times(\mathbf{v}\times\mathbf{B}-\eta\nabla\times\mathbf{B})
\label{eqn:induction}
\end{equation}
Here, $Q$ is the volumetric heating function, $\mathcal{D}$ is the viscous stress tensor, and $\Phi$ represents the viscous heating, which are defined as

\begin{equation}
\mathcal{D}_{ij}=2\bar{\rho}\nu[e_{ij}-\frac{1}{3}(\nabla\cdot\mathbf{v})]\;.
\end{equation} 
\begin{equation}
\Phi=2\bar{\rho}\nu[e_{ij}e_{ij}-\frac{1}{3}(\nabla\cdot\mathbf{v})^2]\;,
\end{equation} 
with $e_{ij}$ as the strain rate tensor. Additionally, both the mass flux and magnetic field are divergenceless: $\nabla\cdot(\bar{\rho}\mathbf{v})=\nabla\cdot\mathbf{B}=0\;.$ Closure is achieved with a linearized equation of state, 

\begin{equation}
\frac{P}{\bar{P}}=\frac{\rho}{\bar{\rho}}+\frac{T}{\bar{T}}\;.
\end{equation}

The calculations were performed within a radial hydrodynamic background state derived using the stellar evolution community code MESA \citep{MESA}. We consider a ZAMS star of 0.4 M$_\odot$ with solar metalicity, a luminosity of $9.478\times10^{31}$ erg s$^{-1}$ ($0.025 L_\odot$), and rotating at twice the solar rate, $\Omega_*=2\Omega_\odot=818\,\mathrm{nHz}$, corresponding to a rotation period of 13.9 days.

 The three simulations presented here (models A, B, and C) are highlights drawn from a far broader survey of the parameter space of dynamo action in these stars, extending our previous work in BT20, and to be published in its entirety in due course. The computational domain of each model extended from just beneath the photosphere, $R_o=0.92R_*=2.382\cdot 10^{10}\,\mathrm{cm}$, to deep in the radiative interior, $R_i=0.1R_*=2.588\cdot 10^{9}\,\mathrm{cm}$. After allowing for thermal equilibration, the base of the convection zone (CZ) was located at approximately $R_{bcz}=0.46R_*=1.19\cdot 10^{10}\,\mathrm{cm}$, resulting in a density stratification in the CZ of roughly $N_\rho=3.5$ scale heights. 

The boundary conditions and diffusion profiles of the models presented here are identical to those employed in BT20, to which we refer the reader for greater detail. In short, the boundaries are impenetrable to convection, stress-free, transmit thermal energy at a rate fixed to balance a volumetric heating term, and require magnetic fields to match onto external potential fields. Diffusion profiles are proportional to $\bar{\rho}^{-0.5}$ in the CZ, and plunge to a floor value in the RZ, with Prandtl number $\mathrm{P_r}=\nu/\kappa=0.25$. 

The three models, A, B, and C, are differentiated by an increasing magnetic Prandtl number $\mathrm{P_m}=\nu/\eta=0.5$, $1$, and $2$, respectively, which predisposes the latter cases to more vigorous dynamo action in the CZ. A hydrodynamical model was evolved first, in the absence of magnetism, to study its general properties. After a statistically steady state was achieved, magnetism was introduced as white-noise perturbations and allowed to self-consistently reshape the flows while growing to its mature amplitudes.

\section{Characteristics of Flows Achieved}

We begin with an examination of the flows realized in our three selected MHD simulations, exploring the patterns of convection established and how they interact with their corresponding magnetic fields. Unless otherwise noted, time averages of hydrodynamical properties are calculated over the full mature time available in each simulation.

\subsection{Longitudinally Modulated Convection}
\begin{figure*}
	\centering
	\includegraphics[width=1.0\linewidth]{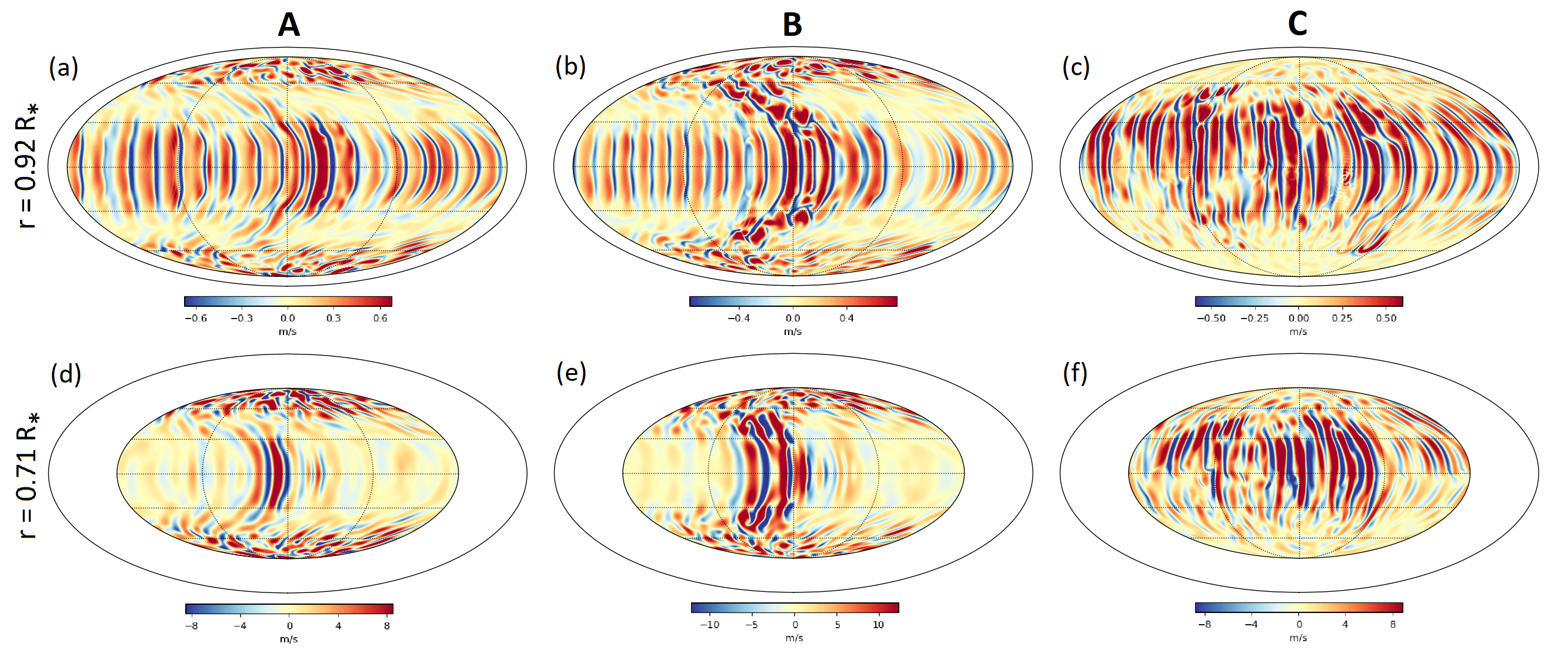}
	\caption{Representative radial velocity $v_r$ in models A, B, and C shown in Mollweide projection, capturing full spherical shells at fixed time, with North and South poles at the top and bottom. Flows are shown both near the surface and near the middle of the CZ, and are rotated to place the striking nests of vigorous convection in the center of the frame. For each model, convection appears as axially-aligned Busse columns near the equator, and as more isotropic cells at high-latitudes in models A and B. In each case, the amplitude of the equatorial flows is strongly modulated in longitude. Near the surface, latitudinal bands of reduced convection form at the edges of the equatorial rolls, which translate to lower latitudes as depth increases, highlighting the active convective nest deeper in the CZ. In model B, the edges of the nest span the latitudinal bands at all depths. Hemispheric asymmetry is evident in model C, with convective flows being partially suppressed by magnetic fields in the southern hemisphere. The longitudinal extent of the nest increases from model A to B, and again from model B to C.}
	\label{fig:convection}
\end{figure*}

\begin{figure*}
	\centering
	\includegraphics[width=1.0\linewidth]{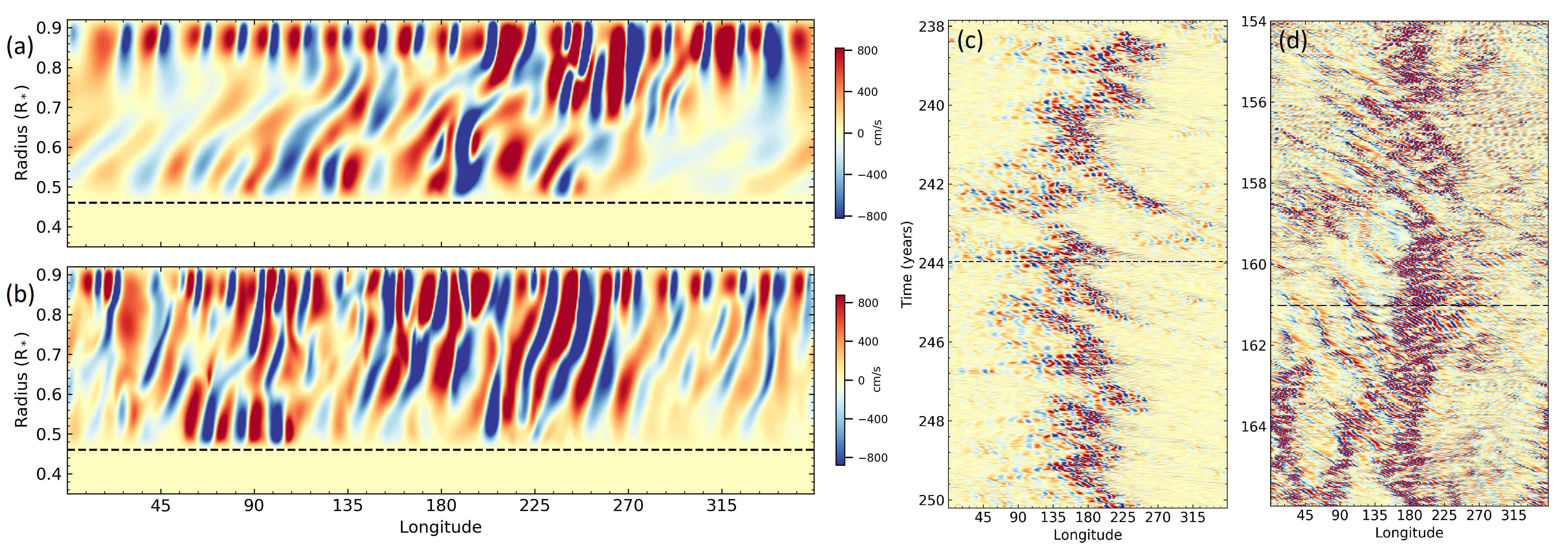}
	\caption{Representations of the radial velocity field $v_r$ in models B and C, showing its structure with depth and evolution in time. (a) An equatorial slice of $v_r$ in model B, unwrapped into a rectangular projection, showing the connectivity of the flows with depth at time $t=243.9$ years. (b) An equatorial slice of $v_r$ in model C at time $t=161.0$ years. (c) Equatorial $v_r$ at depth $r=0.71R_*$ in model B presented in time-longitude space, with a prograde tracking rate of $1.68\cdot10^{-7}$ rad/s $=0.0323\Omega_*$. The nest persists for the entirety of the simulation, propagating prograde relative to the bulk rotation rate and reforming when disrupted. (d) Equatorial $v_r$ at depth $r=0.71R_*$ in model C presented in time-longitude space, with a prograde tracking rate of $2.50\cdot10^{-8}$ rad/s. Relative to model B, there is more significant convective activity outside the nest. Toward the end of the shown time, a secondary nest can be seen forming while the primary nest splits apart.}
	\label{fig:patch_flows}
\end{figure*}

A sample of the convective flows achieved in our three models, A, B, and C, is presented in Figure \ref{fig:convection} as snapshots of the radial velocity $v_r$ near the surface and near mid-CZ. As is common for global convection simulations under significant rotational constraint, we observe that flows near the equator form elongated, axially-aligned rolls known as Busse columns. At higher latitudes in models A and B, these convective rolls are nearly normal to the spherical surface, and thus appear more isotropic. Strong poloidal magnetic fields inhibit the formation of these cells in model C. An exclusionary band of reduced convective amplitude can be seen near the surface at latitudes of approximately $\pm 35^\circ$ in models A and B. These bands translate to lower latitudes deeper in the CZ, before merging with their counterparts from the opposite hemisphere at the equator.

Strikingly, an active \it nest \rm of strong convection is apparent in each model at low latitudes and for all depths in the CZ. The scale of these structures fluctuates over time, and spans roughly $45-90^\circ$ in longitude in models A and B, but is more extended in model C, where it occupies roughly $180^\circ$.  Additionally, the flows of model C demonstrate significant hemispheric asymmetry, with a greater reduction of $v_r$ amplitudes in the southern hemisphere than the north, even affecting the active nest. This asymmetry in the flows is a result of feedbacks from a strong, asymmetric dynamo operating in model C, discussed further in Section \ref{sec:magnetism}.

An equatorial slice of $v_r$ in model B is presented in Figure \ref{fig:patch_flows}(a), showing more clearly how the structure of the convection varies with depth. Outside of the nest, equatorial convective plumes drop off in amplitude sharply below a depth of approximately $r=0.84R_*$, if they continue at all.  Within the nest, flows span the full height of the CZ. At the instant pictured, the deepest pieces of the plumes have broken off and been displaced retrograde (leftward) relative to the nest by a combination of the local differential rotation and the prograde propagation of the nest itself. A comparable equatorial slice is shown for model C in Figure \ref{fig:patch_flows}(b). The alignment of the flows is much more vertical, reflecting the dramatic reduction of shearing flows in that model.

Presented in Figure \ref{fig:patch_flows}(c) is a time-longitude diagram of equatorial $v_r$ in model B at a depth of $0.71R_*$. The frame of the diagram tracks prograde with an angular velocity of $1.68\cdot10^{-7}$ rad/s, the approximate propagation rate of the nest, which persists with intermittent disruption for the full duration of each simulation. This angular velocity is prograde relative to the local rotation rate at the base of the CZ, but slower than both its constituent Busse columns and the near-surface rotation. Of particular note is that the nest appears to have a secondary mode with an angular velocity of $1.94\cdot10^{-7}$ rad/s available to it, faster than the dominant mode but still slower than the Busse columns. It can be seen to be clearly present around 242 years. This mode can also be seen while the slower mode is dominant as prograde distortions which traverse the nest, but do not continue beyond its extents. Figure \ref{fig:patch_flows}(d) shows a similar time-longitude diagram for model C, tracking at a far slower rate of $2.60\cdot10^{-8}$ rad/s. The strong magnetism in model C also strongly quenched its mean azimuthal flows, which measured along the equator at a depth of $0.8R_*$, had a prograde angular velocity of roughly $3.70\cdot10^{-8}$ rad/s, compared to $2.64\cdot10^{-7}$ rad/s in model B. Considering the ratios of these quantities, we find that the equatorial rotation is faster in model B by a factor of 7.0, and the nest propagation faster by a factor of 7.4. The uncertainty of the nest propagation rates is sufficient that these ratios are statistically indistinguishable. It seems that in the presence of strong magnetism, the propagation rates of these nests are linearly related to the amplitude of the differential rotation, and experience the same quenching by Maxwell stresses, though measurements of nest propagation rates in more magnetized models would need to be considered before this could be claimed robustly.

\subsection{Mean Flows and Circulations}

\begin{figure}
	\centering
	\includegraphics[width=0.9\linewidth]{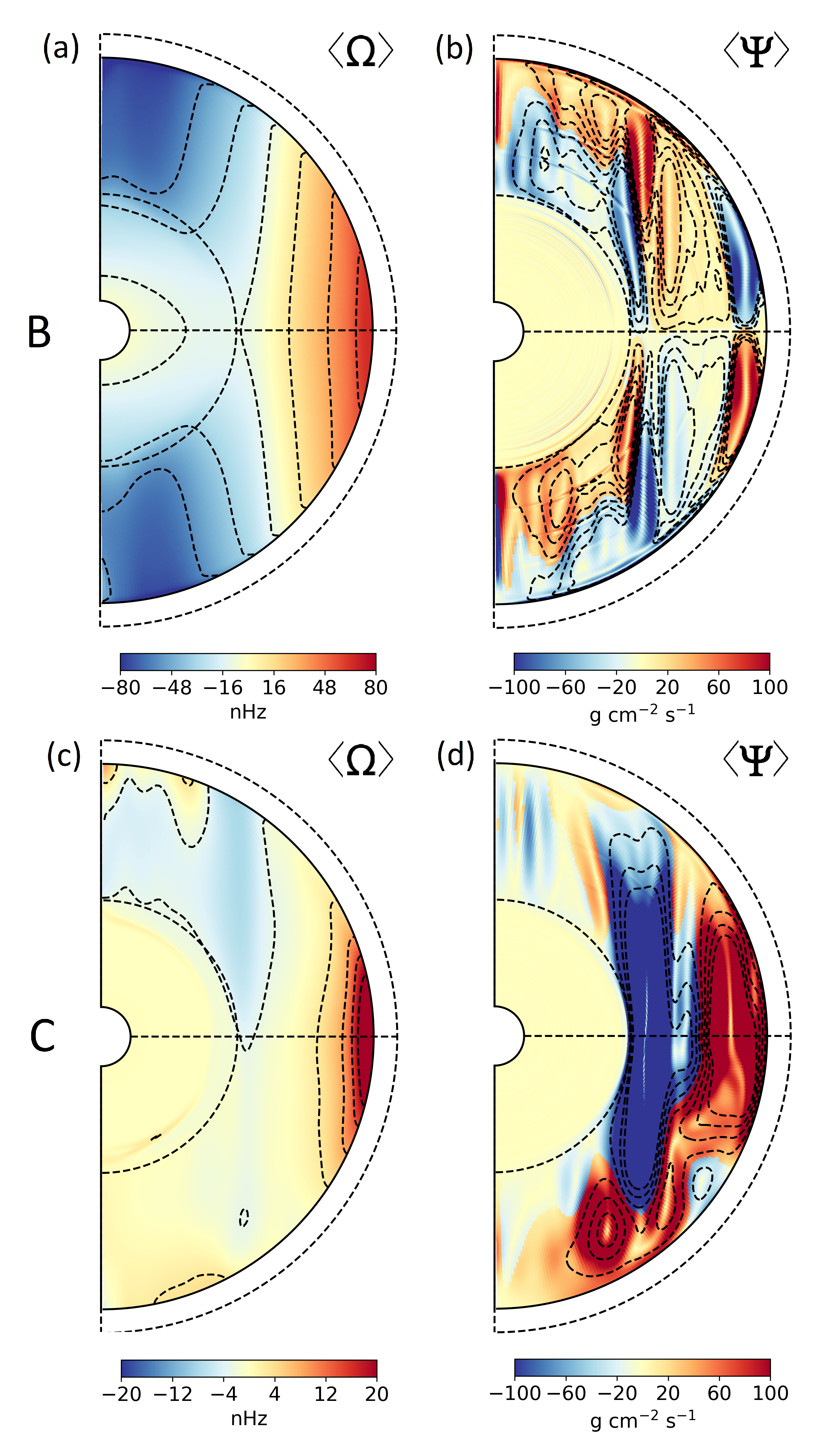}
	\caption{Flows achieved in models B and C, averaged over longitude and time and presented in the meridional plane. (a) Angular velocity in model B is solar-like, with fast equator, slow poles, and a magnetic tachocline. (b) The streamfunction of mass-flux in model B, the contours of which indicate the meridional circulations. Red cells circulate counter-clockwise, and blue cells clockwise. (c) Angular velocity in model C is asymmetric and dramatically reduced in amplitude relative to model B and their hydrodynamic progentior due to the action of strong CZ magnetic fields. (d) Mass flux and meridional ciruclations achieved in model C.}
	\label{fig:mean_flows}
\end{figure}

The differential rotation $\langle\Omega-\Omega_*\rangle_{\phi,t}$ and meridional circulations achieved in models B and C ($\mathrm{Pm}=$ 1 and 2, respectively) is presented in Figure \ref{fig:mean_flows}. Model A ($\mathrm{Pm}=0.5$) is omitted due to its near identical flows to model B. In model B, the differential rotation is solar-like, with fast equators and slow poles, separated by a contrast of $\Delta\Omega = \Omega_{eq}-\Omega_{75} = 149 \mathrm{nHz} = 0.18\Omega_*$. This value matches very closely with that of its hydrodynamical progenitor and of model A. A reasonably thin tachocline appears at the base of the CZ, with a slight negative radial gradient persisting through the RZ to the inner boundary of the model. This gradient is indicative of diffusive imprinting of the CZ flows, which is not wholly curtailed by the magnetic torques in the tachocline. The meridional circulation in model B features thin, equatorward cells at the outer boundary reaching from the equator to about $\pm35^\circ$ in latitude. Notably, the extents of these cells both in latitude and radius corresponds tightly with the position of the exclusionary band of convection. A pair of counter-rotating cells form in each hemisphere at the location of the tangent cylinder, the position in cylindrical radius corresponding to the maximum extents of the stably-stratified RZ. These cells merge onto a distinctly radial two-cell structure in the polar regions, with poleward flows at the edges of the CZ and equatorward flow at mid-depth. Radially thin, stacked circulation cells form in the RZ to maintain a thermal wind balance.

In model C (Figure \ref{fig:mean_flows}c,d), the magnetic influence on the mean flows is much more profound. Largely hemispheric magnetic fields lead to an asymmetric differential rotation profile. While it remains solar-like in the northern hemisphere, though reduced in the amplitude of its contrast to $\Delta\Omega = 31 \mathrm{nHz} = 0.037\Omega_*$, the southern hemisphere is nearly rotating as a solid body. Meridional ciruclations in model C are similarly disrupted by the magnetic fields, and are dominated by a pair of radially stacked cells outside the tangent cylinder. Together, they represent northward axial flows at the tangent cylinder and along the outer boundary, and a southward flow at mid-depth. This structure is further broken up into smaller cells in the proximity of the strong wreaths of toroidal field in the southern hemisphere (see  \S 4).

\subsection{Energy and Momentum Transport}

\begin{figure*}
	\centering
	\includegraphics[width=1.0\linewidth]{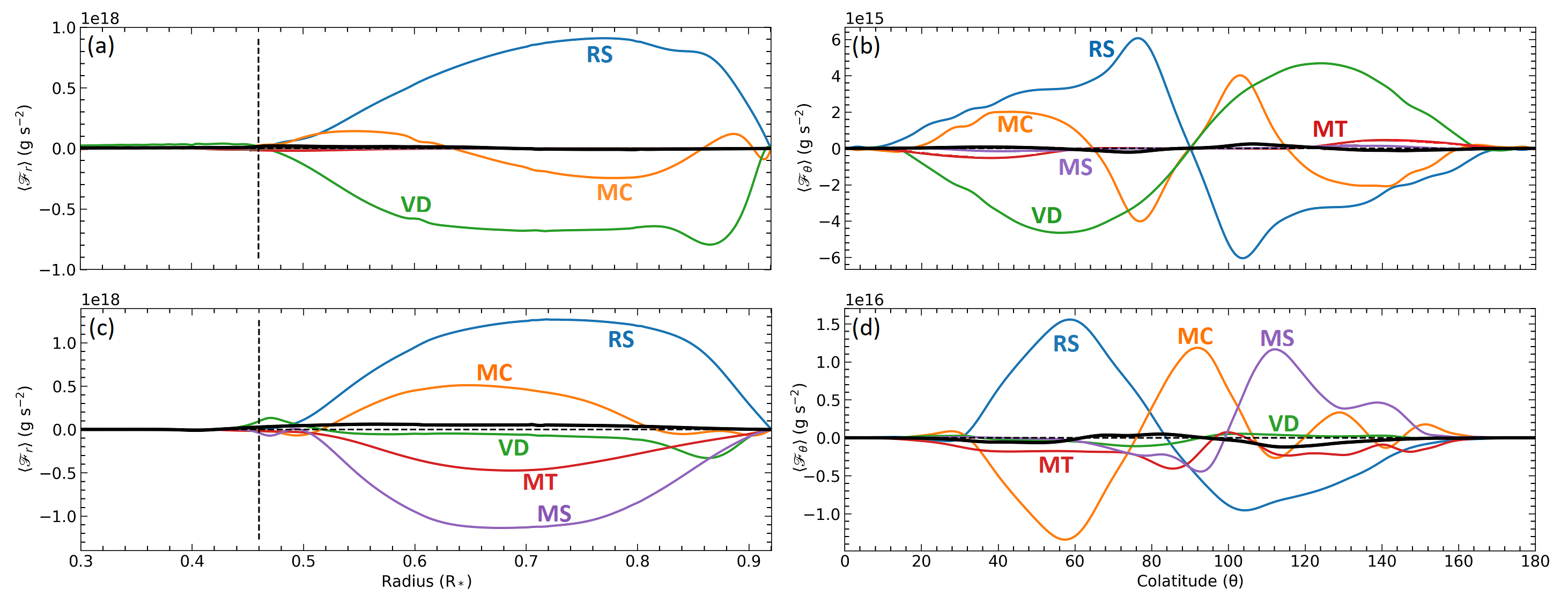}
	\caption{Angular momentum flux balance achieved in models B and C, averaged over longitude and time. (a) Radial flux in model B, additionally averaged over latitude. The Reynolds stress (RS) is balanced primarily by viscosity (VD). (b) Latitudinal flux in model B, additionally averaged over radius. The axisymmetric Lorentz torque (MT) serves to achieve a roughly solid-body interior rotation profile. (c) Radial angular momentum flux in model C. Here, the Lorentz torques (MS, MT) have replaced VD in balancing RS. (d) Latitudinal flux in model C, with the hemispheric asymmetry plainly apparent.}
	\label{fig:amom_flux}
\end{figure*}

\begin{figure}
	\centering
	\includegraphics[width=1\linewidth]{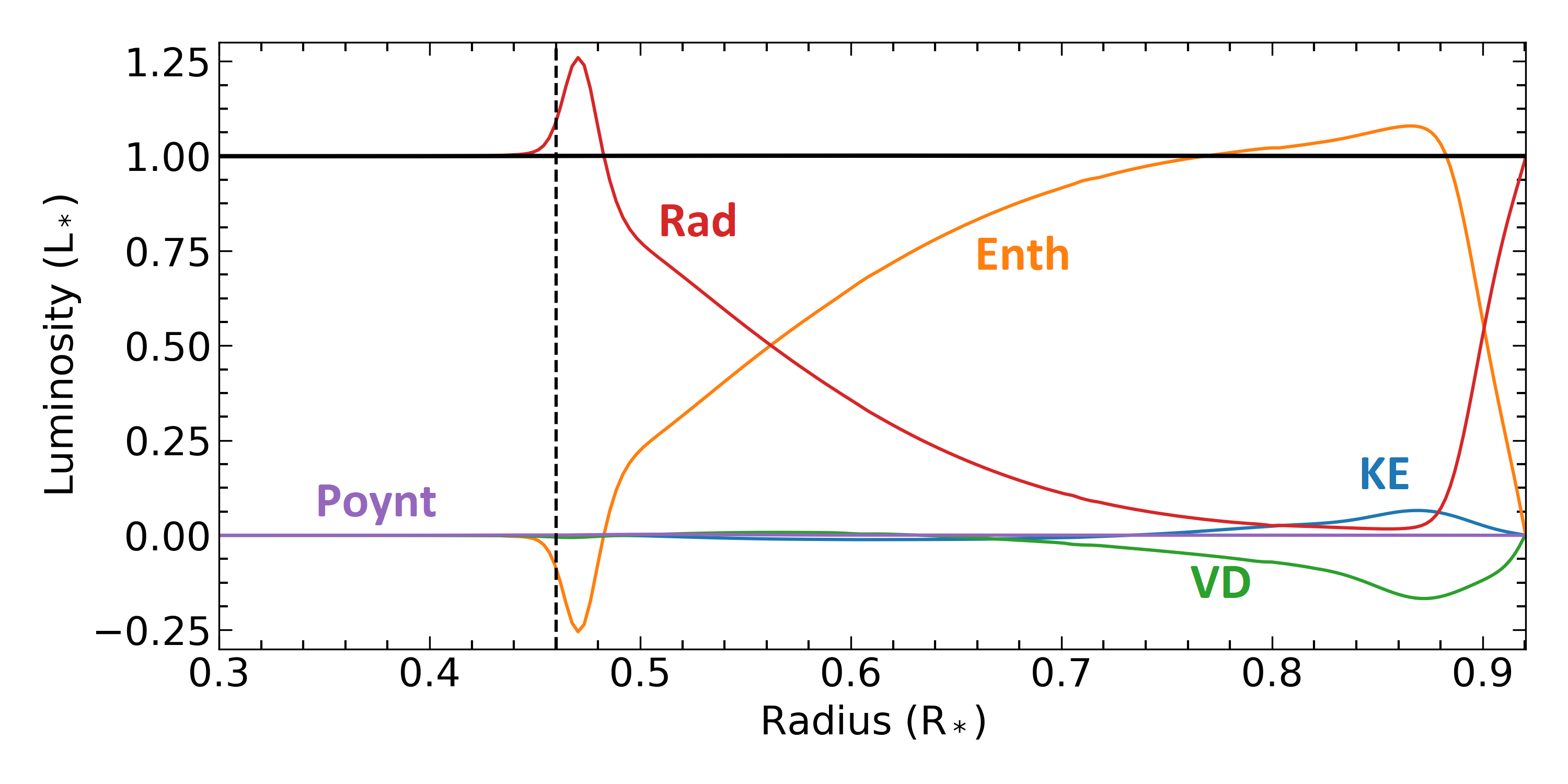}
	\caption{Balance achieved between energy fluxes associated with enthalpy (Enth), kinetic energy (KE), viscous diffusion (VD), Poynting flux (Poynt), and unresolved thermal transport (Rad) in model B, averaged over time and spherical shells. Due to the very small role of the Poynting flux in the overall balance, energy fluxes in model C appear largely the same.}
	\label{fig:energy_flux}
\end{figure}

The flows achieved in our simulations transport energy and angular momentum throughout the computational domain. As these are conserved quantities, in a statistically steady state, the various fluxes of these quantities should come to a balance. In our models, the shell- and time-averaged radial energy flux can be broken up into six terms corresponding to kinetic energy, enthalpy, Poynting, viscosity, conduction, and radiative fluxes, the definitions of which can be found in BT20.
The balance established among these energy fluxes is presented for model B in Figure \ref{fig:energy_flux}. Ascending through the CZ, the radiative flux drops off and is mostly picked up by the enthalpy flux, until reaching the outer thermal boundary layer. A mildly super-luminous enthalpy flux and outward KE flux in the upper reaches of the CZ are balanced by an inward viscous energy flux there. At the base of the CZ, the sign of the mean radial entropy gradient flips, causing the braking of convective plumes and a reversal of the sign of the enthalpy flux. The sharp entropy gradient here also leads to a compensatory spike of outward conductive heat flux, balancing the negative enthalpy. The Poynting flux does not contribute significantly to the overall energy balance in any of our three models. 

The radial and latitudinal angular momentum flux balances are presented for cases B and C in Figure \ref{fig:amom_flux}, averaged in time and longitude. The component fluxes which make up this balance can be attributed to the Reynolds stress (RS), meridional circulations (MC), viscous diffusion (VD), and axisymmetric and non-axisymmetric Maxwell stresses (MT; MS). Again, mathematical definitions of these fluxes are presented in BT20. 

In models A and B, the balance in the CZ is dominated by an opposition between the Reynolds stress and viscous stress. The sign of the transport by meridional circulation can be seen to reverse near the tangent cylinder, reflecting the separation of circulation cells we observe there. The contributions by the magnetic terms are only significant within the tachocline, where they provide poleward fluxes, driving the interior toward solid-body rotation.

In line with the high degree of asymmetry observed with the mean flows of model C, its angular momentum fluxes are also quite skewed. Instead of viscous diffusion, the Reynolds stress in this model is balanced primarily by the non-axisymmetric component of the Maxwell stresses. This balance is typical of models which have quenched a large fraction of their differential rotation, as model C has. With very little differential rotation to imprint, and an increased magnetic opposition to that same rotation, the interior is nearly solid-body.
\section{Characteristics of Magnetism Achieved}
\label{sec:magnetism}

\begin{figure*}
	\centering
	\includegraphics[width=1.0\linewidth]{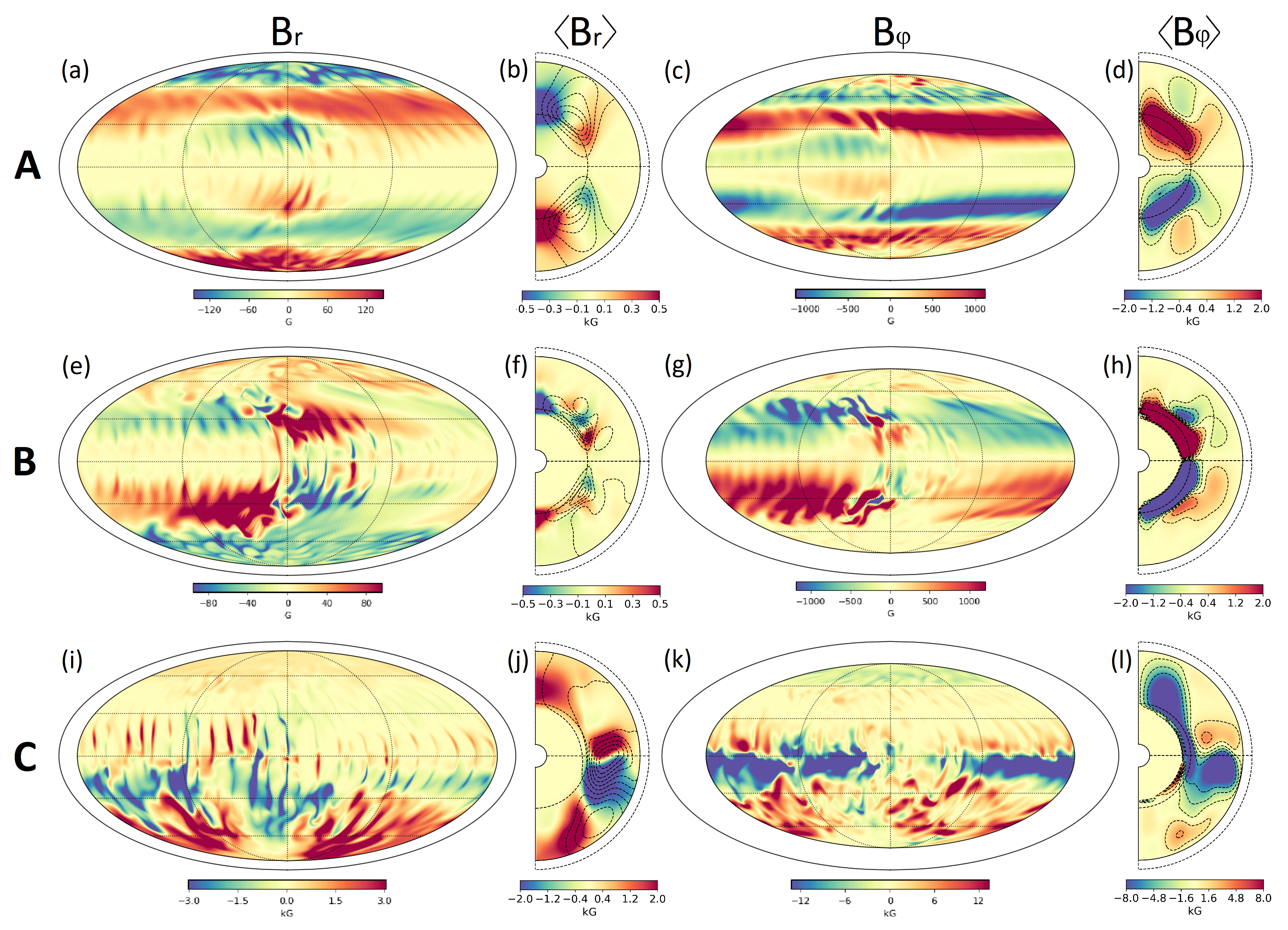}
	\caption{Magnetic structures achieved in models A, B, and C, separated by rows. Radial magnetic fields $B_r$ are shown both in near-surface Mollweide projections at the same times as those in Figure \ref{fig:convection} (a,e,i) and in time- and longitude-averaged meridional projections $\langle B_r \rangle$ (b,f,j). The latter are overplotted with contours of the poloidal streamfunction. Azimuthal magnetic fields $B_\phi$ are shown in Mollweide projections near mid-CZ at $r=0.81R_*$ (c,g,k) and in time- and longitude-averaged meridional projections $\langle B_\phi \rangle$ (d,h,l). For meridional-plane averages of models A and B, oversaturated color bars were chosen to capture CZ field structure; mean magnetic fields in the tachoclines of these models reach $\langle B_r \rangle \sim 1$ kG and $\langle B_\phi \rangle \sim 15-20$ kG. Instantaneous Mollweide projections are rotated such that convective nests align with the center of the frame.}
	\label{fig:mag_structure}
\end{figure*}

We turn next to the magnetic fields built by dynamo action in each of our three MHD simulations. Unless otherwise noted, magnetic quantities are averaged over a single half-cycle of the global dynamo if applicable, or over the same interval as the hydrodynamic variables if the fields are steady.

We present in Figure \ref{fig:mag_structure} representative snapshots and averages of the radial and azimuthal magnetic fields, $B_r$ and $B_\phi$ respectively, for each of our three MHD models. The mean fields of model A are steady in time, with powerful, antisymmetric wreaths of toroidal field filling the tachocline and reaching amplitudes of roughly 18 kG. Through the CZ, the fields of model A diminish in amplitude but remain largely time-steady and axisymmetric, splitting into two wreaths in each hemisphere, antisymmetric across the equator, with amplitudes around 1 kG near mid-depth at $r=0.81R_*$. The poloidal fields of model A are dominated by their axisymmetric dipole and octupole moments, achieving peak values of about 3 kG in the tachocline and diminishing in amplitude as they approach the stellar surface. For both $B_r$ and $B_\phi$, the influence of the equatorial band of longitudinally modulated convection is apparent. Throughout the CZ, there are virtually no magnetic fields at the equator except within the traveling nest, where they possess the opposite sense and comparable poloidal amplitudes to the nearest wreath..

The magnetic fields of model B resemble those of A within the tachocline, again filling each hemisphere with strong, antisymmetric, time-steady fields reaching approximately 20 kG. Through the CZ, however, the character of the magnetic fields achieved in model B is quite different to those of A. Here, the azimuthal fields again have typical amplitudes around 1 kG, but they form only a single wreath in each hemisphere which extends very near to the equator. Again, the position of the traveling nest can be identified by the presence of magnetic fields with a reversed sense relative to the mean, but in this case the reversal extends to nearly all latitudes with strong $B_\phi$. The poloidal fields induced at the tachocline in model B are quite similar to those in model A, with dominant axisymmetric dipolar and octupolar modes, but do not extend through the CZ as cleanly, instead breaking up into higher order, largely non-axisymmetric structures. 

\begin{figure*}
	\centering
	\includegraphics[width=1.0\linewidth]{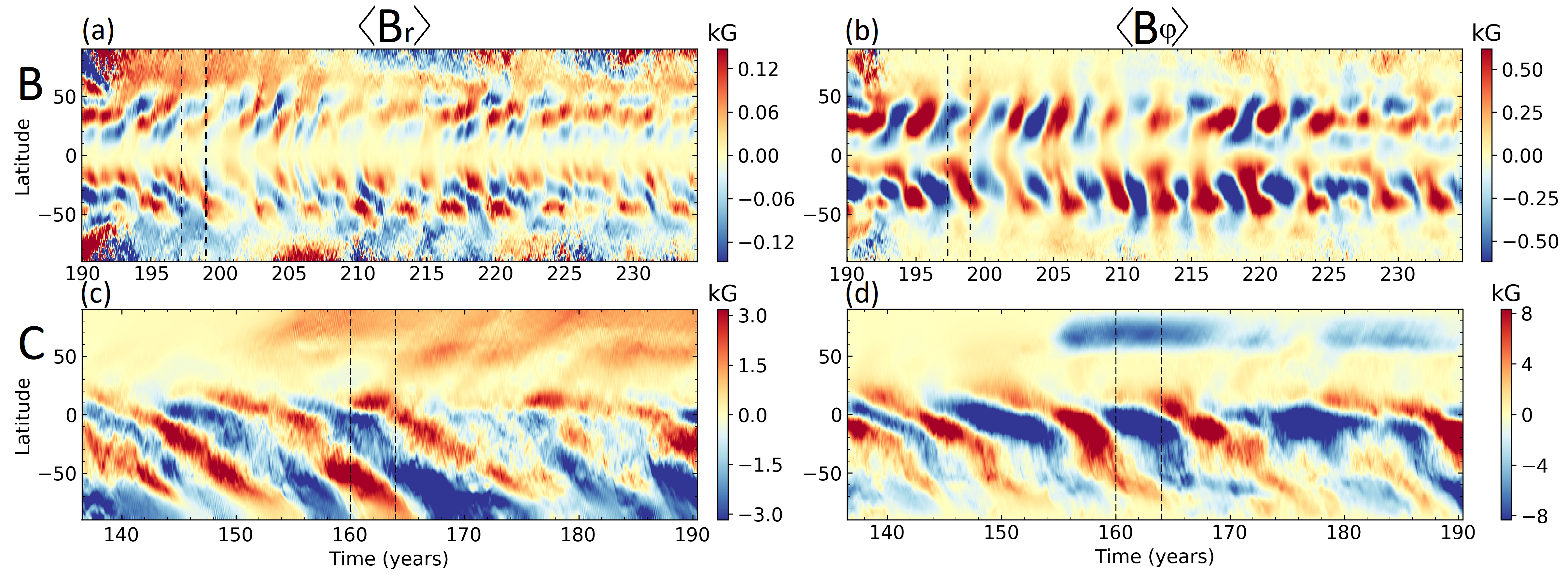}
	\caption{Time-latitude diagrams showing the evolution of longitude-averaged magnetic fields in models B and C. Model A is not included due to its mean fields being steady in time. Both fields are shown near the middle of the CZ at depth $r=0.74R_*$. Vertical dashed lines indicate the intervals over which the meridional projections in Figure \ref{fig:mag_structure} were averaged. While both models B and C undergo cycles in their CZs, the timing of the cycles in model B is much faster and much more regular.}
	\label{fig:mag_evolution}
\end{figure*}

While the tachocline fields of model B are steady in time, the fields in its CZ undergo regular cycles with a period of $T_B=1.75$ years. Time-longitude diagrams of mid-CZ $\langle B_r \rangle$ and $\langle B_\phi \rangle$, akin to solar butterfly diagrams, for models B and C are presented in Figure \ref{fig:mag_evolution}. At the instant selected, the CZ fields of model B are antisymmetric across the equator, but the two hemispheres are not strongly bound. Over the course of its evolution, slight phasing differences in each hemisphere accumulate, leading model B to wander between symmetric and antisymmetric states. The magnetic fields at latitudes $|\theta|>60^\circ$ do not conform to the cycles occurring at lower latitudes, and undergo an irregular cycle with an average reversal time of about 5 years. $\langle B_r \rangle$ here is comparable in amplitude to the low-latitude fields, but $\langle B_\phi \rangle$ is significantly lower, around 100 G.

Model C achieved and maintained a particularly novel configuration for its magnetic fields. Unlike models A and B, the CZ here produces magnetic fields of comparable amplitude to those in its tachocline, with $B_\phi$ peaking around 20 kG in each. The fields thus produced were largely restricted to the southern hemisphere, at times reaching across the equator to about $\theta=20^\circ$ in latitude. The greater longitudinal extents of the nest, coupled with the stronger toroidal fields in model C lead to a more complex interaction than in models A or B. As in B, we see greatly enhanced poloidal fields above and in the wake of the nest, but the amplitude of the reversed toroidal field at the core of the nest is typically far less than what exists outside it. In the northern hemisphere, fields from the time-steady tachocline imprinted unobstructed through the CZ, resulting in large-scale axisymmetric fields that show little evolution in time. In the southern hemisphere, however, the cycling, highly structured magnetic fields in the CZ instead imprinted into the tachocline. These fields propagate poleward throughout the course of their 11 year cycle, though a break can be seen in their propagation rate as they traverse the tangent cylinder around $\theta=-50^\circ$.

\subsection{Nest-Mediated Dynamo Cycles}
\begin{figure}
	\centering
	\includegraphics[width=1\linewidth]{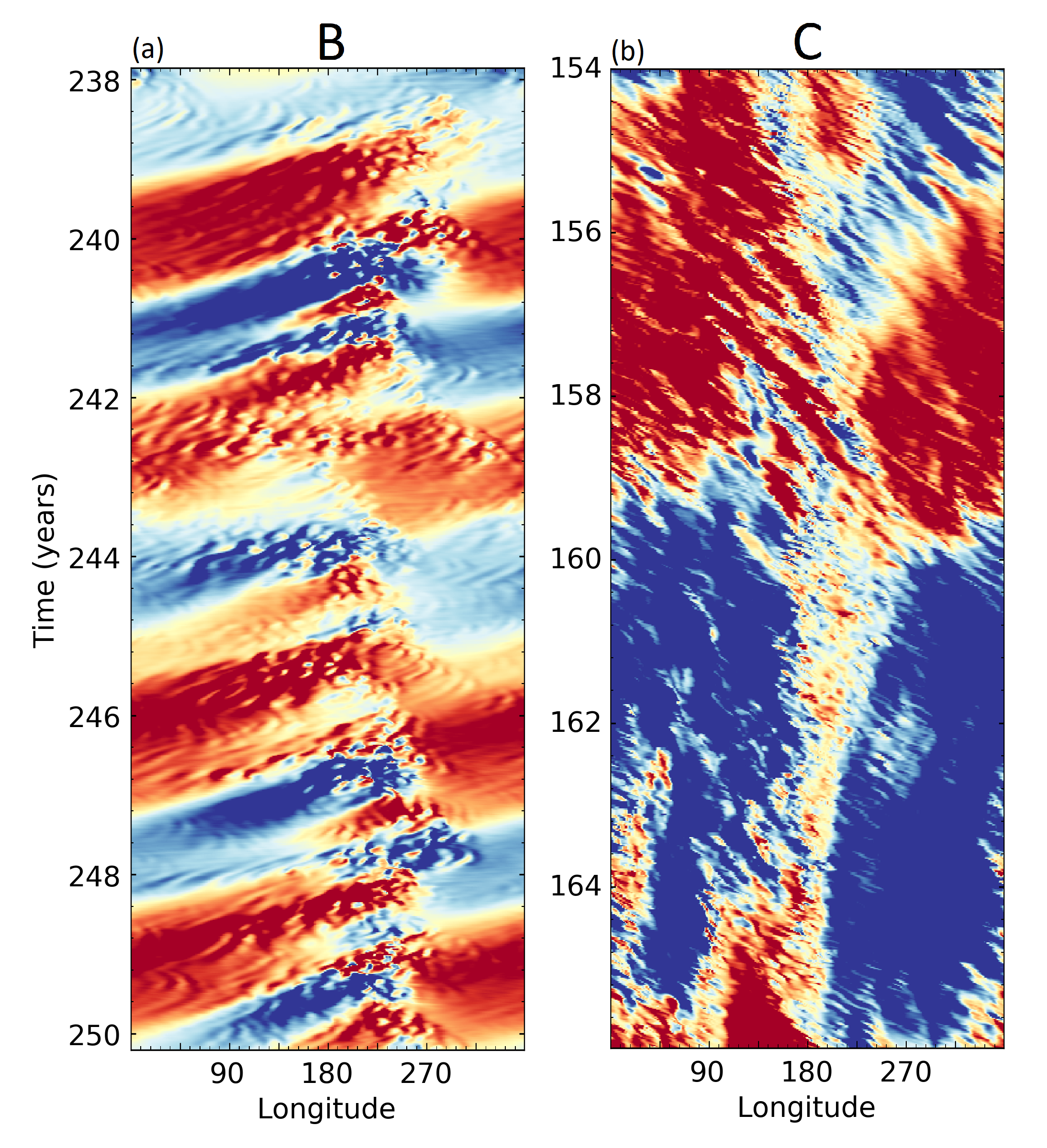}
	\caption{$B_\phi$ at depth $r=0.81R_*$ and a latitude of $-30^\circ$ presented in time-longitude space, and tracked prograde at the propagation rate of the convective nest for models B and C (B: $1.68\cdot10^{-7}$ rad/s, C: $2.50\cdot10^{-8}$ rad/s). In model B, fields propagate slower than the convective nest, and thus appear to travel down and to the left. In model C, magnetic fields are more closely tied to Busse columns, which propagate prograde more rapidly through the patch, rightward in this frame. Upon reaching the leading edge of the nest, the sign of $B_\phi$ tends to reverse. After $t=162$ years, a secondary nest begins forming, which also reverses $B_\phi$.}
	\label{fig:patch_fields}
\end{figure}

Because the magnetic fields in the CZ are built principally through a combination of differential rotation and helical convection, the consolidation of low-latitude convective vigor into a traveling nest presents a migratory, localized source of turbulent induction in these stars. In Figure \ref{fig:patch_fields}, we present time-longitude diagrams of $B_\phi$ in models B and C, again tracking at a rate which fixes the traveling nest near the centerline of each panel. From it, we can observe a clear difference in the ways that the magnetic fields of these two models interact with the nest. 

\begin{figure*}
	\centering
	\includegraphics[width=0.9\linewidth]{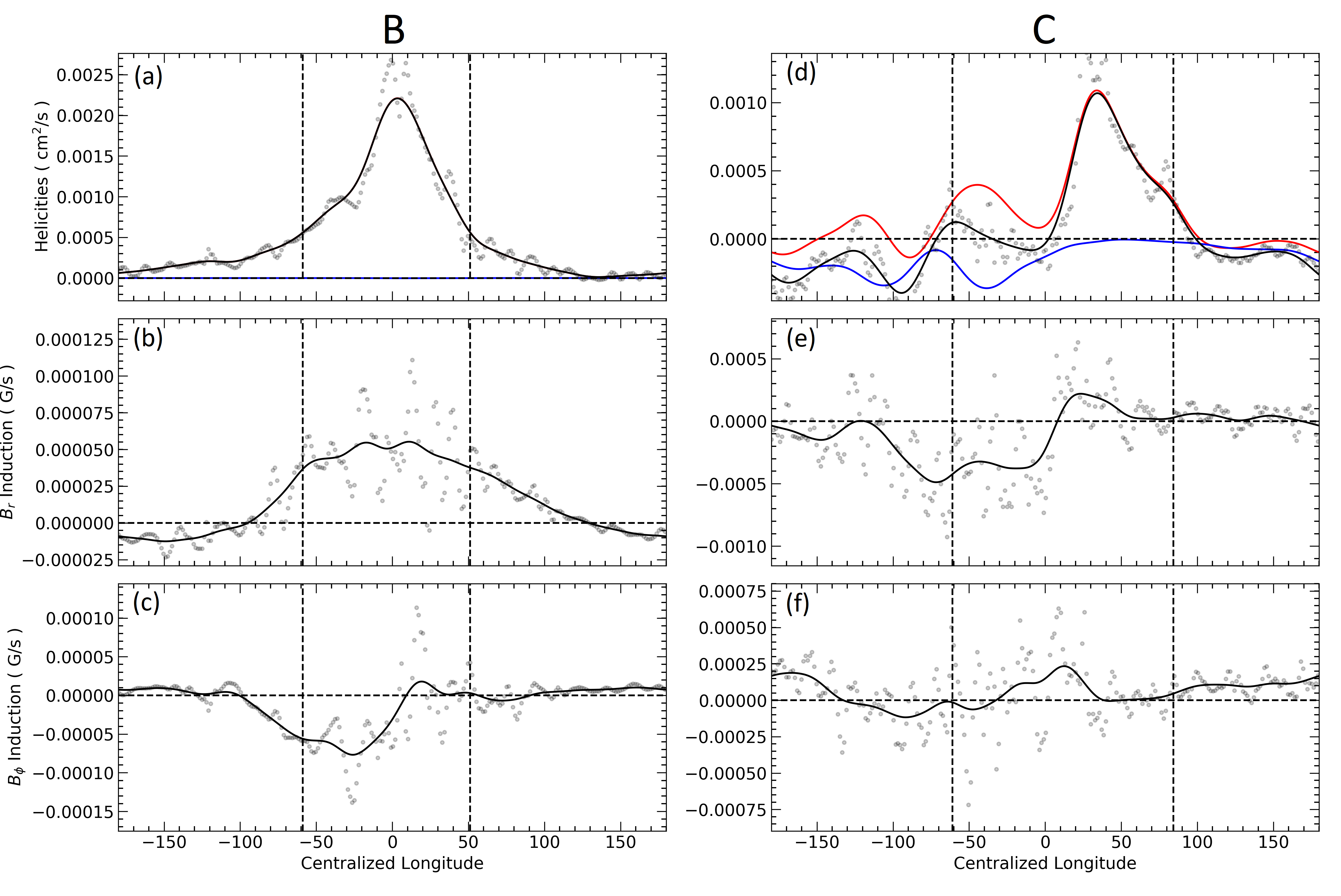}
	\caption{Comparison of inductive processes in models B and C at a depth $0.81R_*$, averaged from $-45^\circ$ to $0^\circ$ in latitude, and averaged in time after realigning with a central longitude for the convective nest. Convective noise is smoothed with a gaussian filter. The boundaries of the nests, estimated by the FWQM of the flow helicities, are plotted as vertical dashed lines. (a) Total helicity in model B. The kinetic helicity in the nest dominates the total, resulting in a sharply peaked $\alpha$-effect. (b) Shear-driven radial induction $S_r$ for model B. Poloidal field generation is positive within the nest, and mildly negative outside it. (c) Shear-driven $\phi$ induction $S_\phi$ in model B. Toroidal fields are generated primarily within the nest and in its wake. (d) Kinetic (red), current (blue), and net helicities (black) in model C. The nest shows a strong peak of kinetic helicity at its leading edge, but strong current helicities saturate the tail half of the nest and drive a negative $\alpha$-effect outside it. (e) $S_r$ for model C, with its greatest amplitudes appearing within and in the tail of the nest. The sign reverses relative to the mean in the leading edge of the nest (f) $S_\phi$ for model C, showing a sharp positive peak in the leading edge of the nest and a negative peak at its rear.}
	\label{fig:patch_induction}
\end{figure*}

In model B, we observe a retrograde propagation (down and to the left) of $B_\phi$ structures relative to the nest. As these structures complete a lap of the equator and arrive at the leading edge of the nest, we see that they are reversed in sign, which is maintained as they exit the trailing edge of the nest and begin their next lap of the star. If we instead consider the interaction from a frame corotating with the magnetic fields, the image becomes a bit like that of a snake eating its own tail. The traveling nest, the mouth of the snake, chews up magnetic fields left behind by its previous passage, the tail, reversing them. The propagation rate of the magnetic fields varies somewhat from cycle to cycle, but averages $2.20\cdot10^{-8}$ rad/s, which is very similar to the Alfv\'en speed of the fields produced in the nest. Comparing that to the angular velocity of the nest at $1.68\cdot10^{-7}$ rad/s, we are able to compute a beat period of 1.36 years. This sets a floor for the reversal time in model B, pending the inductive timescale for reversing the fields once a lap is complete. Considering an average toroidal induction rate in the nest of $\dot{B_\phi} = 5\cdot10^{-5}$ G/s and a canonical tail amplitude of $B_\phi=500$ G, we estimate this timescale to be $\tau_B = B_\phi / \dot{B_\phi} = 0.32$ years. When combined with the lapping time, we estimate a reversal time of 1.68 years, which almost perfectly recovers the observed reversal time of 1.75 years, which was calculated independently through Fourier analysis of the mean magnetic fields. As such, we are confident that these interactions with the nest are the primary driver for the global reversals observed in model B. 

The fields of model C also exhibit a reversal of their sign upon entering the nest, however it does not lead to global reversals. In the wake of its passing, magnetic fields revert to match the sign of the pre-existing mean fields, which are much stronger than those of model B. When these mean fields do eventually reverse, they begin growing first within and in the wake of the nest.

Why does the convective nest cause the fields it interacts with to reverse in sign? In descriptions of stellar dynamos (e.g. \citealt{parker55}; \citealt{pouquet76}; \citealt{moffatt78}), inductive processes proportional to the curl of the magnetic field are known as $\alpha$-effects, and are primarily responsible for the production of poloidal field from toroidal. In models of stellar CZs, the most prominent $\alpha$-effect is caused by helical convective motions. It is proportional to the kinetic helicity $H_k$ of the fluid, and confounded by current helicities $H_c$ which emerge in opposition as a saturation mechanism at high amplitudes of the magnetic field:

\begin{equation}
\alpha_0 = \frac{-\tau_c}{3}(\underbrace{\langle\mathbf{v}\cdot\nabla\times\mathbf{v}\rangle}_{H_k} + \underbrace{\frac{-1}{4\pi\bar{\rho}}\langle \mathbf{B}\cdot\nabla\times\mathbf{B} \rangle}_{H_c})\;,
\end{equation}

with $\tau_c$ the auto-correlation time of convective flows. If we consider the distributions of kinetic and current helicity presented in Figure \ref{fig:patch_induction}(a) for model B, we can see that $\alpha_0$ clearly changes character within the nest. $H_k$ peaks strongly there in response to the enhanced convective amplitudes, and dominates the $\alpha$-effect. Figure \ref{fig:patch_induction}(b) shows the turbulent shear-induction rate of $B_r$ in model B, defined as

\begin{equation}
S_r = [\mathbf{B}\cdot\nabla\mathbf{v}]_r \approx \alpha_0[\nabla\times \mathbf{B}]_r\;.
\end{equation}

We can see that the shear-generation of $B_r$ in the CZ of model B is negative outside of the nest, but positive and of greater amplitude within it. In solar-like dynamo models, the regeneration of toroidal fields from poloidal is dominated by the large-scale shear of differential rotation, the $\Omega$-effect, though $\alpha$-effects may play a significant or even dominant role in some regimes. Figure \ref{fig:patch_induction}(c) shows the combined shear-induction rate of $B_\phi$ in model B, defined as

\begin{equation}
S_\phi = [\mathbf{B}\cdot\nabla\mathbf{v}]_\phi \approx \alpha_0[\nabla\times \mathbf{B}]_\phi+(\mathbf{B_m}\cdot\nabla)[\Omega r\mathrm{sin}\theta]\;,
\end{equation}

where $\mathbf{B_m}=B_r\hat{r}+B_\theta\hat{\theta}$. As with the poloidal field, we can see that the majority of $B_\phi$ induction in model B occurs within and in the immediate wake of the nest. Due to the time it takes for $B_r$ to build up, this happens past the peak longitude for the $\alpha$-effect, and so it can easily be attributed to the $\Omega$-effect. The alignment of these two effects is such that the $B_\phi$ produced there is of the opposite sense to what arrives at the leading edge. 

Considering the helicities of model C, shown in Figure \ref{fig:patch_induction}(d), the leading edge of the nest maintains a recognizable peak of $H_k$, but the overall flow helicity is significantly quenched in the trailing half. Moreover, the strong magnetic fields of model C generate a current helicity which saturates the $\alpha$-effect there, and drives it to negative values outside of the nest. The radial field induction rate, shown in Figure \ref{fig:patch_induction}(e) reflects this difference in nest characteristics, with a small positive peak near the leading edge which dives to a strong negative peak at the tail of the nest. As with the radial fields, Figure \ref{fig:patch_induction}(f) shows that model C has a slight bias toward positive $B_\phi$ generation near the front of the nest, and a strong negative peak at its tail. Model C differs from model B in this regard, in that it does not maintain any substantial differential rotation in the southern hemisphere, precluding an $\Omega$-effect. The shear generation rates of $B_r$ and $B_\phi$ follow the same general pattern within the nest in model C, indicating that both processes are dominated by the $\alpha$-effect there. The reversal of the sign of the $\alpha$-effect in the tail of the nest suggests that the magnetic fields there are being returned to the configuration they entered with. The magnetic energy transferred from the toroidal fields to the poloidal within the nest is returned to the toroidal fields. Due to the overall amplification provided by the $\alpha$-effect, however, the energy is returned to the toroidal fields in the wake of the nest with interest, leading $B_\phi$ to attain its maximal values there. 

The shear profile of model C is not \it completely \rm dominated by the convective cells, however. Although magnetic torques prevent the convection from establishing large-scale differential rotation in the southern hemisphere, the peaked $H_k$ at the leading edge of the nest is indicative of a localized deviation from that trend. The enhanced $\alpha$-effect provided by the leading edge of the nest diminishes $B_\phi$ there enough that a modest radial contrast of $\Delta\Omega_r =12.7$ nHz can be maintained at the equator within the nest, as opposed to $\Delta\Omega_r =3.2$ nHz outside it. This contrast, measured at the equator from $r=0.85R_*$ to $r=0.6R_*$, provides a non-axisymmetric $\Omega$-effect in the leading edge of the nest, which uses the locally-induced poloidal fields to reverse the sense of $B_\phi$ there. Because the processes maintaining this magnetic structure in model C would be quenched by strong fields, the reversed $B_\phi$ produced by the nest is fixed at relatively low amplitude and cannot lead to global reversals as it does in model B, instead remaining localized within the nest.  

\begin{figure*}
	\centering
	\includegraphics[width=0.9\linewidth]{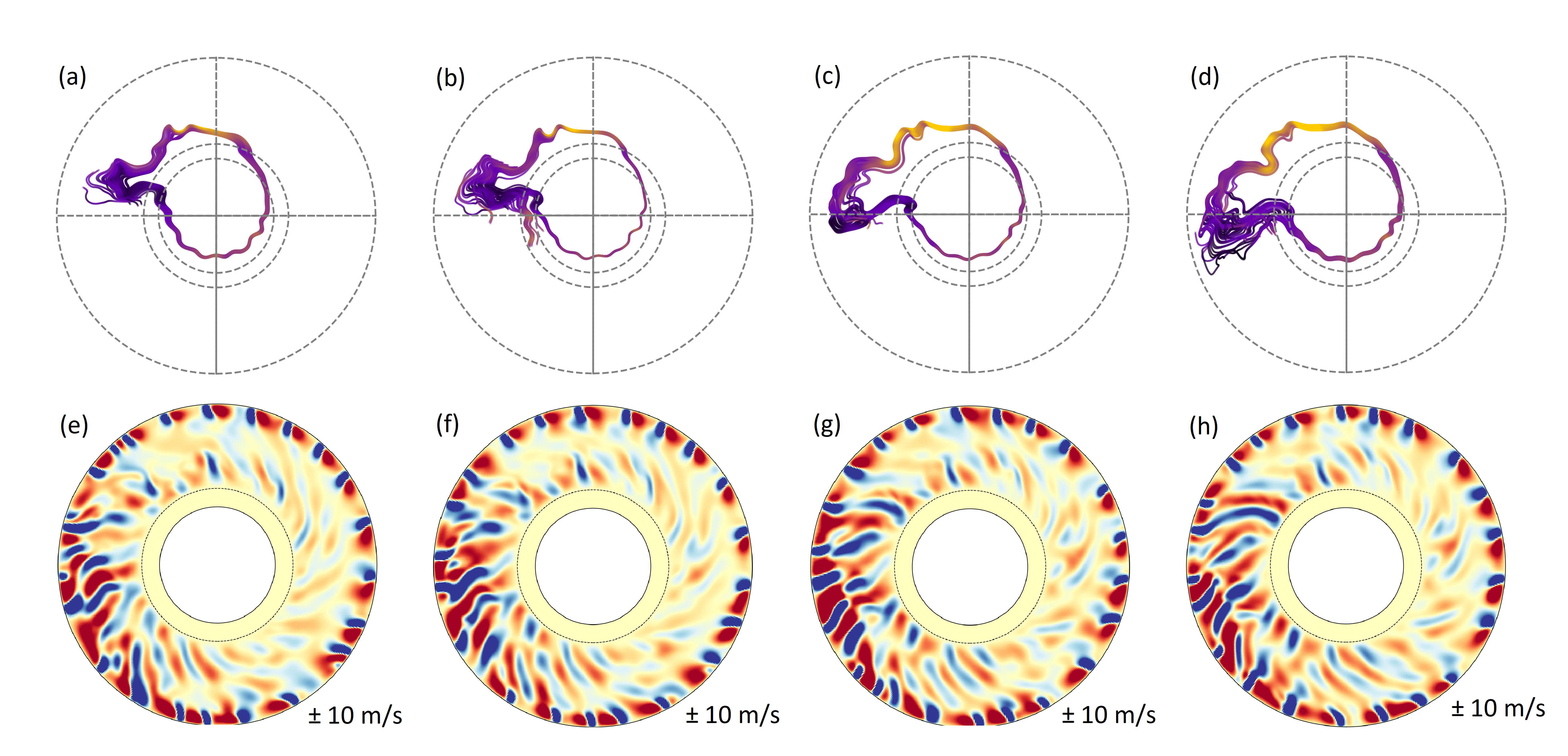}
	\caption{(a-d) A sequence of field-line tracings of a rising magnetic loop identified in case D2ta of BT20, shown from above the north pole. Each frame is separated by an interval of roughly 6 days. Brighter colors represent stronger fields, ranging from 2 kG to 30 kG. (e-h) Equatorial slices of $v_r$ at times corresponding to the above frames. A nest of enhanced convection coincides with the rising flux rope.}
	\label{fig:rising_loops}
\end{figure*}

Despite their proximity in parameter space, models B and C present very different versions of the same star. Which, if either, should be held as a canonical representation of how longitudinally modulated convection impacts dynamo action? It is tempting to claim that model C, with its more vigorous and turbulent magnetic field generation, should be more indicative of a real M-dwarf interior, and that the nest-driven cycles of model B are little more than a curiosity. However, the strong fields of model C nearly entirely eliminate the differential rotation in the CZ, which does not appear to be a universal feature of magnetically active stars, but certainly impacts the inductive balances in our models. In that respect, models A and B, with their strong differential rotation may be more realistic representations of some stars. Ultimately, true stars likely feature elements of both models, along with other behaviors that may not be captured here.

\subsection{Longitudinally Modulated Flux-Emergence}

In concluding their hydrodynamical discussion of convective nests, \citet{brown08} proposed that these structures may in turn lead to the formation of magnetic active longitudes, which have been observed on the Sun and other cool stars for decades. In all three of the models presented here, we find that the enhanced kinetic helicity and vertical advection provided by an active nest leads to the star's near-surface poloidal fields being greatest in amplitude in and around the nest. Furthermore, in the work of \citet{nelson13}, it was shown that buoyant magnetic flux ropes tend to form in localized regions of intense $B_\phi$, and that their rises are often dictated as much by convective motions as by their own magnetic buoyancy. Thus, the pattern we see in all of our models with $B_\phi$ peaking just on the tail of the convective nest suggests that the trailing edges of these structures may be ideal places to build magnetic flux ropes and transport them to the surface where they can become starspots. 

In BT20, we reported on a set of M-dwarf simulations some of which were quite similar to those reported here. In particular, model D2ta from that work was identical to model B here, except in the boundaries of its computational domain and its treatment of the tachocline. Model D2ta had a slightly taller CZ, reaching nearer to the stellar surface and capturing $N_\rho=5$ density scale-heights, and a shallower RZ, which terminated at $r_i=0.35R_*$. Though it was nominally more akin to model B, its increased stratification resulted in a greater magnetic Reynolds number $\mathrm{R_m} = v_{rms}L/\eta$, which in turn led to a dynamo configuration bearing similarities to all three models reported here. It had extremely strong, time-steady magnetic fields in both the CZ and tachocline, reaching mean field strengths on the order of 30 kG and peaks which in places exceeded 80 kG. The convection of model D2ta followed the same patterns of longitudinal modulation identified here, which led also to the same localized reversals of $B_\phi$ and concentrations of the poloidal field. 

In subsequent analysis of the magnetic fields in model D2ta, however, we also found that it possessed a number of self-consistently formed magnetic flux ropes rising from its tachocline. A series of field line tracings for one such flux rope is shown in Figure \ref{fig:rising_loops}, along with equatorial slices of $v_r$ at the same times. The diffusivities in the CZ of model D2ta are not particularly low, and so any compact, high-amplitude magnetic structures it forms tend to resistively leak away their magnetic energy before the associated buoyancy can carry them very far. This can be seen already in panel (a), where $|B|$ drops to around 10 kG in the crest of the characteristic $\Omega$-shaped loop nearly as soon as it is formed. Due to the loop's positioning on the trailing edge of the convective nest, however, it is able to straddle a strong upflow and ride it all the way to the outer boundary of the computational domain. Convective nests can promote the longitudinally-localized rise of magnetic flux ropes not just with their upflows, but also through the effect they have on $B_\phi$ in their immediate wake. In panels (c) and (d), another loop can be seen taking shape in the upper left quadrant, where field strengths exceed 30 kG. The $B_\phi$ attains its maximal values here, and thus we can expect that the formation of buoyant proto-loops should occur with a higher frequency. Considering their effects on the magnetic fields in and around them, both at small and large scales, nests of enhanced convection could reasonably contribute to the formation of active longitudes on the Sun and other stars. We reserve further discussion of the flux ropes identified in model D2ta, as well as the machine learning tool we developed for the purpose, for an upcoming paper. 

\section{Conclusions}

We have presented the results of three global MHD simulations of fast-rotating M2-like stars with tachoclines. Each of these three models, A, B, and C featured a longitudinally-modulated nest of enhanced convection which propagated prograde along the equator through a band of otherwise diminished convective amplitude. We observed that in the presence of strong magnetism, the propagation rates of these nests remained proportional to the differential rotation measured at the equator.  

We explored in detail the interactions of this convective nest with the magnetic fields generated in each model, finding that the strong $\alpha$-effect it provides tends to result in enhanced poloidal fields and reversed toroidal fields within it. As in model B, this effect can become the dominant inductive process in the stellar CZ and set the cycling period of the global fields. Additionally, we found that the trailing edge of a convective nest can become a preferential site for flux-emergence, both in terms of large-scale poloidal fields and for magnetic flux ropes. We suggest that if present in real stars, these traveling nests of convection may contribute to the formation of persistent active longitudes that have been observed on the Sun and other stars.

The potential for these travelling convective nests to produce persistent, observable defects on the surfaces of stars presents a problem for observational determinations of stellar rotation rates. Where spectroscopic measurements are unavailable, observers often turn to periodic fluctuations in a star's brightness, attributed to surface defects such as starspots rotating in and out of view. As we have shown here, travelling convective nests appear to be a preferential site for flux emergence. If the resulting starspots remain bound to and travel with the convective nest, a behavior which our models are unable to test, then the prograde propagation of these structures may lead to a systematic overestimation of rotation rates in the stars which build them. This would amount to an error on the order of a few percent for fast-rotating stars with photometric rotation rates.

In the three models presented here, we observed a wide range of magnetic behaviors across a very narrow region of parameter space. Varying only $\mathrm{P_m}$, we observed a globally steady interface dynamo, an interface dynamo coupled to a regularly cycling CZ, and a hybrid state with a steady interface dynamo in one hemisphere and a cycling distributed dynamo in the other, all while maintaining a longitudinally modulated nest of enhanced convection. It is clear that dynamo action in the deep CZs of M-dwarfs can be fascinatingly sensitive and diverse, and the work presented here still leaves many fundamental questions unanswered. How are these small, dim stars able to produce the spectacular magnetic activity they have become known for? How might the processes we have observed change in FC stars without tachoclines? We look forward to further exploration and endeavour toward understanding these commonplace yet mysterious stars.

\software
{MESA \citep{mesasw},
Rayleigh \citep{rayleighsw}}

\acknowledgements
{We thank Brad Hindman and Loren Matilsky for helpful conversations in the development of this work. We thank Nick Featherstone for authoring the Rayleigh code, as well as the Computational Infrastructure for Geodynamics, which is funded by the National Science Foundation. Computational resources for this project were provided by the NASA High End Computing (HEC) program through the Pleiades supercomputer at NASA Advanced Supercomputing (NAS) in the Ames Research Center, as well as local computational infrastructure. This work was supported by NASA Astrophysical Theory Program grant NNX17AG22G and FINESST grant 80NSSC20K1543, as well as Heliophysics grant 80NSSC18K1127 for partial shared support of the computational infrastructure needed to analyze the major dynamo simulations.}

\end{document}